\documentclass[fleqn,usenatbib]{mnras}
\usepackage{newtxtext,newtxmath}
\usepackage[T1]{fontenc}
\usepackage{ae,aecompl}
\usepackage{footmisc}
\PassOptionsToPackage{pdfpagelabels=true}{hyperref}
\usepackage{graphicx}	% Including figure files
\usepackage{amsmath}	% Advanced maths commands
\usepackage{amssymb}	% Extra maths symbols
\usepackage{threeparttable, booktabs}
\usepackage{ulem}
\graphicspath{{/Users/annachilds/M6/plotsforPaper1/}}
\hypersetup{final}

\title[Giant planet effects on terrestrial planet formation and system architecture]{Giant planet effects on terrestrial planet formation and system architecture}

\author[Childs et al.]{
Anna C. Childs,$^{1}$
Elisa Quintana,$^{2}$
Thomas Barclay,$^{2,3}$ and
Jason H. Steffen$^{1, \thanks{E-mail: jason.steffen@unlv.edu}}$
\\
$^{1}$Department of Physics and Astronomy, University of Nevada, Las Vegas, NV 89154, USA\\
$^{2}$NASA Goddard Space Flight Center, 8800 Greenbelt Road, Greenbelt, MD, USA\\
$^{3}$University of Maryland, Baltimore County, 1000 Hilltop Cir, Baltimore, MD 21250, USA\\
}

\date{Accepted XXX. Received YYY; in original form ZZZ}

\pubyear{2017}

\begin{document}
\label{firstpage}
\pagerange{\pageref{firstpage}--\pageref{lastpage}}
\maketitle

% Abstract of the paper
\begin{abstract}
Using an updated collision model, we conduct a suite of high resolution N-body integrations to probe the relationship between giant planet mass, and terrestrial planet formation and system architecture.  We vary the mass of the planets that reside at Jupiter's and Saturn's orbit and examine the effects on the interior terrestrial system.  We find that massive giant planets are more likely to eject material from the outer edge of the terrestrial disk and produce terrestrial planets that are on smaller, more circular orbits.  We do not find a strong correlation between exterior giant planet mass and the number of Earth analogues (analogous in mass and semi-major axis) produced in the system.  These results allow us to make predictions on the nature of terrestrial planets orbiting distant Sun-like star systems that harbor giant planet companions on long orbits---systems which will be a priority for NASA's upcoming Wide-Field Infrared Survey Telescope (WFIRST) mission.

\end{abstract}


\section{Introduction}
Terrestrial planet formation is typically separated into three stages due to the different physical processes that are involved in each period \citep{Lissauer93,Righter65}.  The initial stage involves the condensation of solids from the gas disc that grow until they reach kilometer-size planetesimals \citep{Chiang10}.  The middle stage focuses on the agglomeration of small planetesimals (Moon-sized bodies) into embryos \citep[Mars-sized bodies][]{Weidenschilling77,Rafikov03}.  The final stage follows the collisional evolution of embryos, the giant impact phase, that ultimately yields planets \citep{Chambers01,Agnor99}.  

When modeling the late stage of planet formation, we assume that all the gas in the disc has been dispersed.  The lifetime of a protoplanetary gas disc depends on the dispersal mechanisms of the gas, but for solar-type stars the lifetime of the gas disc is approximately 2-3 Myr \citep{Alexander06,Pfalzner14}.  After the gas disc dissipates, the growth of protoplanets ensues via gravitational collisions that yield accretion and erosion events \citep{Righter11,Wetherill95}.  

A shortcoming of the core-accretion model is its inability to replicate the mass and formation timescale differences between Earth and Mars \citep{Raymond09}. A less conventional picture of terrestrial planet formation which addresses this inability, is the ``Pebble Accretion" model where 100-km to 1000-km bodies accrete submeter-sized pebbles to form terrestrial planets \citep{Lambrechts12,Levison15}.  Most studies that utilize N-body integrators however, model the late stage of planet formation.  The physics involved considers only the interactions of planetesimals and embryos as purely gravitational, and until recently, assumed only relatively trivial collisions---either completely elastic or completely inelastic.  These commonly used collision models limit the accuracy of simulations of terrestrial planet formation, as real collision outcomes are more complex \citep{Chambers13,Haghighipour17}.  An accurate collision model is needed to probe the properties of planetary systems that depend on the collision history, such as their final architecture, composition, magnetic field, moon system, atmosphere and internal structure \citep{Lock17,Elser11,Jacobson17}.

\cite{Leinhardt12} developed a prescription for more realistic collisions that includes fragmentation.  \cite{Chambers13} implemented this fragmentation model into the N-body integrator \textit{Mercury} to allow for collisions that result in partial accretion or erosion.  With this \textit{fragmentation code}, \cite{Quintana16} studied how giant impacts affect terrestrial planet formation in the presence of Jupiter and Saturn.  They found that the final architecture of the terrestrial planetary system is comparable to systems formed using a trivial collision model, but the collision history and accretion timescales of these systems differed significantly---giving insight into a different formation process.  Examining the high-resolution collision history of a planet also has implications for habitability.  Tracking the fragments in a system allows us to examine how the byproducts of collisions interact with the rest of the system and affect planet formation.  Further analysis of these interactions will determine if they result in impacts which are catastrophic to the planet's habitable properties---stripping the atmosphere, heating the planet, fragmenting the planet, etc.

When considering the solar system, not only is a high-resolution collision study needed to accurately explore the properties of the terrestrial planets, but the effects of the giant planets must be accounted for as well.  Previous studies have shown the important role Jupiter played in the formation of our terrestrial system.  \cite{Horner08} showed that Jupiter shielded earth from a high rate of giant impacts from the asteroid belt.  More generally, \cite{Levison03} found that different exterior giant planet environments produce terrestrial planets of different sizes and orbits.  Additionally, \cite{Raymond14} found that changing the eccentricity of Jupiter and Saturn will affect the orbital and mass distributions, and water delivery to the terrestrial planets.

Other studies suggest that exterior giant planets may be necessary for the development of life on Earth analogues \citep{Horner10}.  \cite{Horner09} found through numerical studies that exterior giant planets promote the delivery of volatiles to planets in the habitable zone of the system.  Additionally, \cite{Horner15} argues that giant planets are needed to induce climate change on planets in the habitable zone, similar to the role Jupiter plays in the Milankovitch cycles on Earth.  Considering the influence that the exterior giant planets had on the solar system, the question follows: What would our terrestrial planets be like if Jupiter and Saturn did not grow large enough to accrete significant amounts of gas before the gas disc dissipated, and therefore had much smaller masses?

Along these same lines, thousands of diverse exoplanet systems have been discovered by the \textit{Kepler} mission.  Although \textit{Kepler} is not sensitive to planets on longer orbital periods, \cite{Foreman16} developed an automated method to predict the occurrence rates of giant planets with long orbital periods, even if only one or two transits were observed.  Their probabilistic model estimates the occurrence rate for exoplanets on orbital periods from 2-25 years, with a radii between 0.1-1$R_J$ to be $\sim$ 2.00 per G/K dwarf star.  On the other hand, \cite{Wittenmyer16} found from radial-velocity surveys that $\sim$ 6$\%$ of stars host giant planets (a planet with a minimum mass of 0.3 $M_\text{Jup}$) in orbits between 3 and 7 AU.  Although there is some disagreement on the occurrence rates, future missions will help constrain these rates and it will be useful to understand how giant planets affect terrestrial planet formation.  \cite{Bryan18} looked at systems containing super-Earths and found that $39\pm7 \%$ of these systems host long period companions between 0.5-20 $M_{Jup}$ and 1-20 AU, suggesting that there exists a correlation between the presence of super-Earths and giant planets on long orbits.  \cite{Zhu18} found a stronger correlation between super-Earths and cold Jupiters, predicting $\sim90\%$ of systems with cold Jupiters contain super-Earths, and that cold Jupiters occur around $32 \pm 8$ of Sun-like stars.  Our numerical results presented here show that giant planets affect terrestrial planet formation in a variety of ways.

Upcoming missions are being designed to search for planetary systems that resemble our solar system with a higher sensitivity to planets on long orbits.  For example, NASA's Wide-Field Infrared Survey Telescope (WFIRST) is expected to launch in the mid-2020's.  Some of the mission objectives are to determine how common planetary systems that resemble the solar system are, what types of planets are on longer orbital periods, and what determines the habitability of Earth-like worlds.  This mission aims to resolve the discrepancy on giant planet occurrence rates.

Here we explore the extent to which exterior giant planets affect the formation and evolution, and the final architecture of the terrestrial planet system.  We use a suite of high resolution N-body simulations with five different giant planet environments with giant planets at Jupiter's and Saturn's current orbit.  Each suite of simulations has scaled-down masses for Jupiter and Saturn---maintaining the 3:1 mass ratio between the body at Jupiter's orbit and the body at Saturn's orbit.  We use exterior giant planets that are $\sim$ $\tfrac{3}{4}$,$\tfrac{1}{2}$, $\tfrac{1}{4}$ and $\tfrac{1}{7}$ the mass of Jupiter and Saturn.  Because \cite{Suzuki16} argues that cold Neptunes are likely the most common type of planets beyond the snow line, the smallest mass used in our giant planet systems is a Neptune sized body.  The integration time and the varied masses used for Saturn and Jupiter are listed in Table \ref{tab:ic}.  If we can constrain the role that exterior giant planets play in terrestrial planet formation, we can predict how common solar system-like planetary systems are, in conjunction with the occurrence rates of planets on long orbits from WFIRST.

\section{Simulations}

\begin{table}
\begin{center}
\caption{The exterior giant planet masses and integration times used in our simulations.  For reference, Jupiter is 318 $M_{\oplus}$ and Saturn is 95 $M_{\oplus}$.
}
\label{tab:ic}
\begin{threeparttable}
\renewcommand{\arraystretch}{1.2}% for the vertical padding
\begin{tabular}{ccc}
\toprule
Mass at Saturn's orbit  & Mass at Jupiter's orbit & Time \\($M_{\oplus}$)& ($M_{\oplus}$)& (Myr) \\
\hline
95& 318& 500\\
75& 225& 5\\
50 & 150& 500\\
30& 90& 5\\
15& 45& 5\\

\bottomrule
\end{tabular}
\end{threeparttable}
\end{center}
\end{table}

\begin{figure}
\begin{center}
\includegraphics[width=\columnwidth]{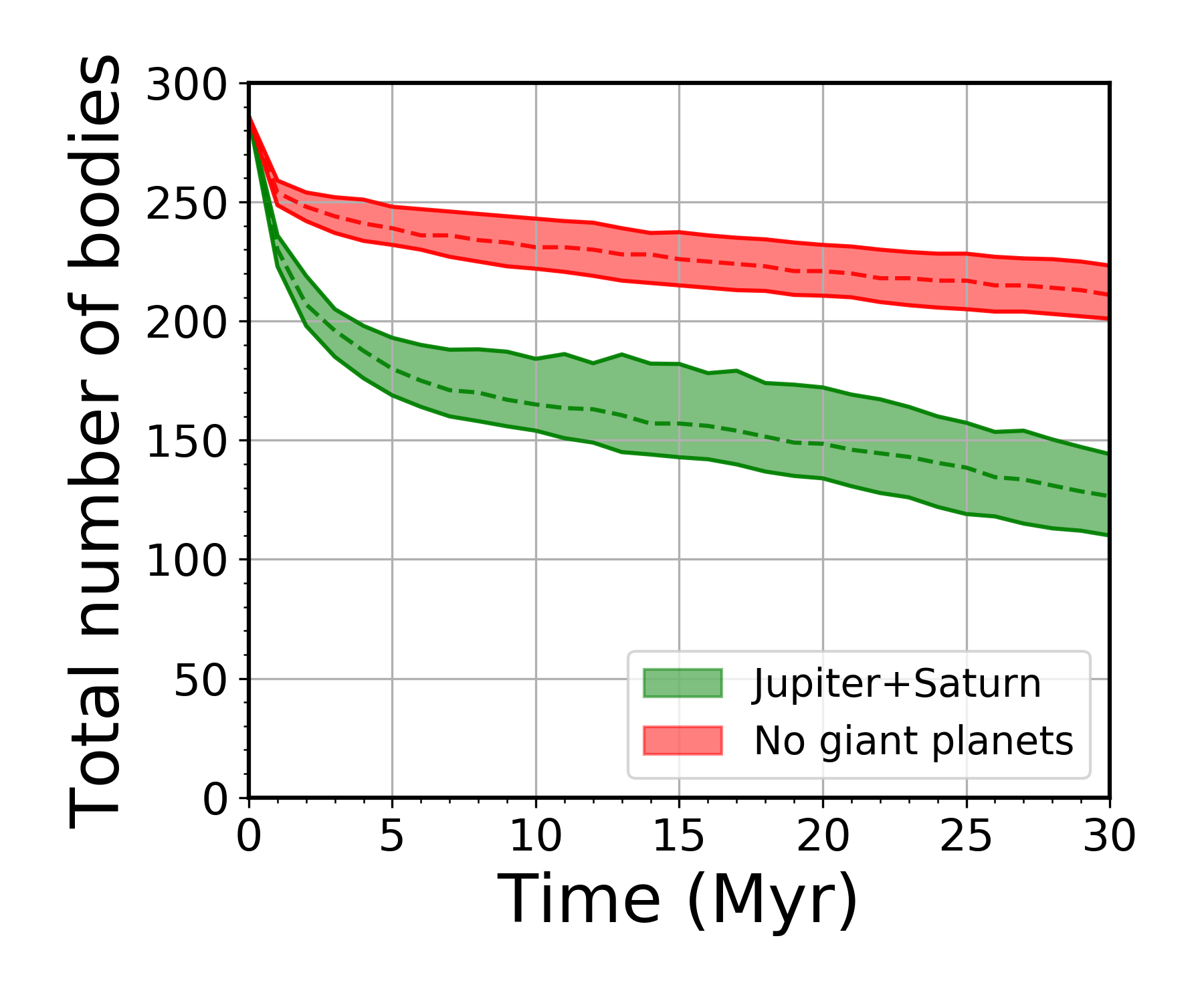}
\caption{Reproduced from \protect\cite{Barclay16}, the total number of bodies (embryos and planetesimals) for 140 simulations done for a system with Jupiter and Saturn at their present orbits, and a system without giant planets.  $1-\sigma$ ($16^\text{th}$ and $84^\text{th}$ percentiles) ranges are the lower and upper bounds, and the dashed line represent the $50^\text{th}$ percentile of the system.}
\label{fig:tom}
\end{center}
\end{figure}

The \textit{fragmentation code} developed by \cite{Chambers13} (adopting the collision model from \cite{Leinhardt12}) allows for several of collision outcomes.  These include:
\begin{itemize}
\item  A collision with the central star if the object comes within 0.1 AU of the star
\item Ejection of a body from the simulation if its semi-major axis becomes larger than 100 AU
\item Completely inelastic collision (merger)
\item A grazing event, which results in partial erosion or accretion after less than half of the impactor interacts with the target
\item Head-on, or super-catastrophic collision, in which one of the bodies loses $\geq$ half of its mass
\item A hit-and-run collision, in which the target mass does not change, but the impactor may result in partial erosion \citep{Genda12}.
\end{itemize}

Since the CPU time scales with the square of the number of bodies in the system a minimum fragment mass must be set in order to reduce the computational expense as the number of fragments grow.  Our minimum fragment mass is set to $m=0.0047M_\oplus$.  Erosive events occur when a body fragments and the excess mass is above or equal to the minimum fragment mass. If the excess mass from an erosive event meets this requirement, the \textit{fragmentation code} will split the excess mass into fragments of equal size.  If this mass requirement is not met, the collision will not result in any mass loss.

Using this code, two extreme cases were studied by \cite{Barclay16}---a system with Jupiter \& Saturn, and a system with no exterior giant planets.  Figure \ref{fig:tom} shows their results for the number of bodies in the system versus simulation time for 140 simulations (along with $16^\text{th}$ and $84^\text{th}$ percentile bounds).  From this plot it is clear that the two tracks can be distinguished within the first 5 Myr.  Consequently, most cases that we study here have a simulation time of 5 Myr.  We integrate two of our simulation suites up to 500 Myr (one with Jupiter \& Saturn and one with planets of mass 150 \& 50 $M_{\oplus}$) so that we may consider the evolution of the system over a longer timescale.  We use a 7 day timestep in our integrations which is $\approx \tfrac{1}{10}$ the time of the innermost orbit period.

Previous N-body studies have used discs of small planetesimals and larger planetary embryos to successfully reproduce the broad characteristics of the solar system's terrestrial planets \citep{Chambers01}.  As a result, similar mass distributions are commonly used in protoplanetary discs for N-body simulations studying terrestrial planet formation around Sun-like stars.  The disc used in our simulations was adopted from \cite{Quintana14}, which was an extrapolation of the disc used by \cite{Chambers01}.  Our disc contains 26 embryos (Mars-sized, $r = 0.56 R_\oplus$; $m=0.093 M_\oplus$), and 260 planetesimals (Moon-sized, $r=0.26 R_\oplus$; $m=0.0093 M_\oplus$) yielding a total disc mass of 4.85 $M_{\oplus}$.  The disc has no gas.  This bimodal mass distribution marks the epoch of planet formation which is dominated by purely gravitational collisions (the late-stage) \citep{Kokubo00}.  All masses have a uniform density of 3 g cm$^{-3}$.  The surface density distribution $\Sigma$ of the planetesimals and embryos follows $\Sigma \sim r^{-3/2}$, the estimated surface density distribution of Solar Nebula models \citep{Weidenschilling77}.  These masses are distributed between 0.35 AU and 4 AU from a solar-mass star.

The eccentricities and inclinations for each body are drawn from a uniform distribution with $e< 0.01$ and $i< 1^{\circ}$.  The argument of periastron, mean anomaly, and longitude of ascending node are chosen at random.  Exterior planets with different masses are placed at Saturn's and Jupiter's orbit, 5.2 AU and 9.6 AU respectively, with their present orbital elements.  We ran a suite of simulations for each scenario---each comprising 150 simulations with a slight change of one planetesimal's longitude of ascending node.  Chaotic evolution takes these small changes in one planetesimal and rapidly produces entirely different systems.  Our main emphasis in this work is to determine the effects of a varying exterior planet mass on the mass that remains in the regions where terrestrial planets eventually form.

\section{Results}

\begin{figure}
\begin{center}

\includegraphics[width=\columnwidth]{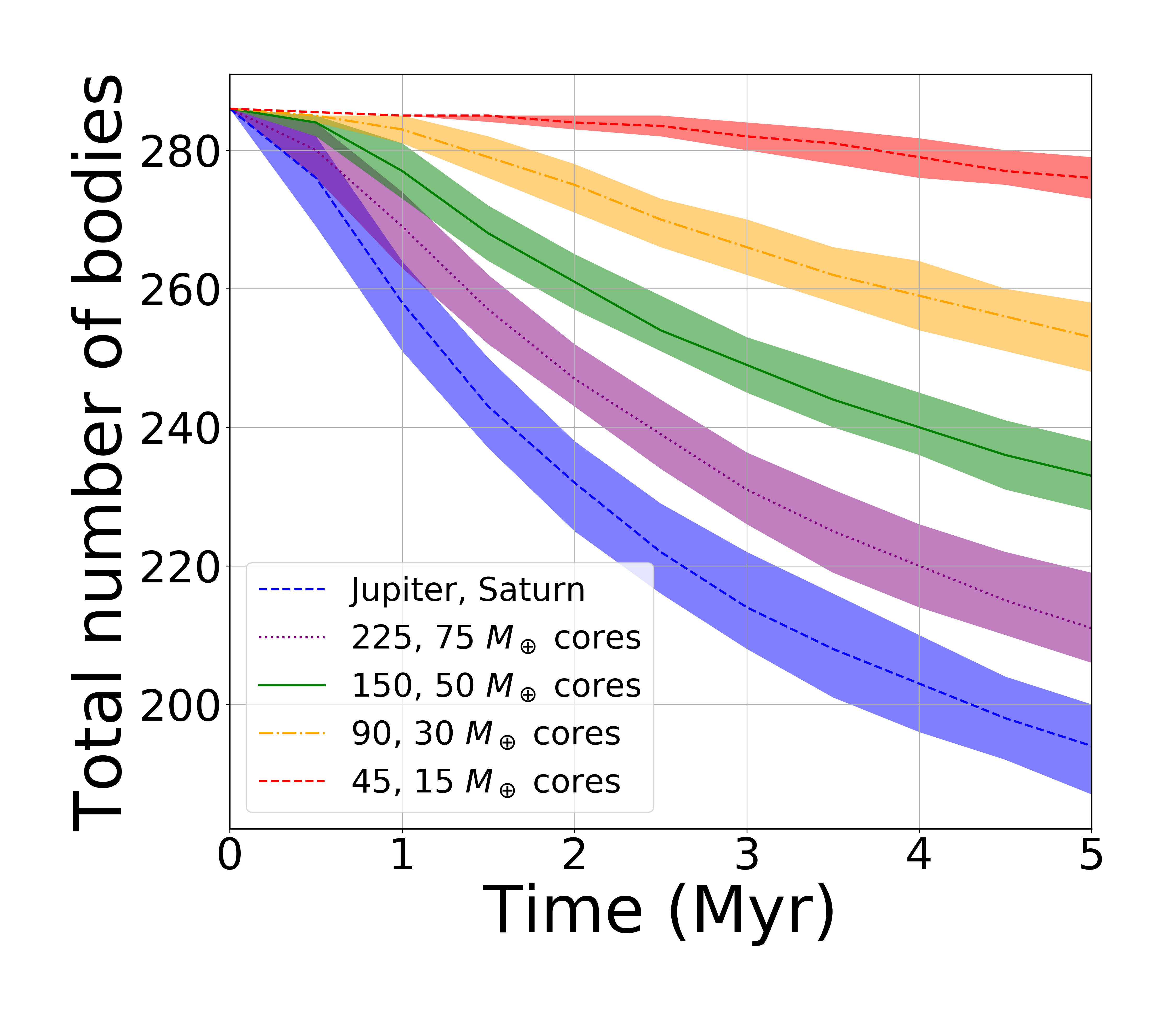}
\caption{The total number of embryos and planetesimals versus integration time for all 150 simulations done for each system.  $1-\sigma$ bounds ($16^\text{th}$ and $84^\text{th}$ percentiles) are shaded and the medians are the respective center lines.}
\label{fig:bvt}
\end{center}
\end{figure}

\begin{table}
\begin{center}
\caption{The average time to eject 10\% of the bodies (planetesimals and embryos) from the system compared to the average time for Jupiter \& Saturn to eject 10\% of bodies in the system.%  The total, exterior giant planet mass of the system is compared to the mass of Jupiter+Saturn.
}
\label{tab:reduction_rates}
\begin{threeparttable}
\renewcommand{\arraystretch}{1.2}% for the vertical padding
\begin{tabular}{cccc}
\toprule
Exterior masses & Exterior mass ratio & Time & Time ratio \\ ($M_{\oplus}$) & & (Myr) \\
\hline

Saturn \& Jupiter & 1 & 0.5 $\pm$ 0.1  & 1 \\
75 \& 225& 0.73 & 1.2 $\pm$ 0.2 & 2.4 \\
50 \& 150& 0.48 & 1.9 $\pm$ 0.3 & 3.8 \\
30 \& 90& 0.29 & 3.7 $\pm$ 0.6 & 7.4\\
15 \& 45 & 0.15 & $>$ 5 & $>$ 10 \\

\bottomrule
\end{tabular}
\end{threeparttable}
\end{center}
\end{table}

\begin{figure}
\begin{center}
\includegraphics[width=\columnwidth]{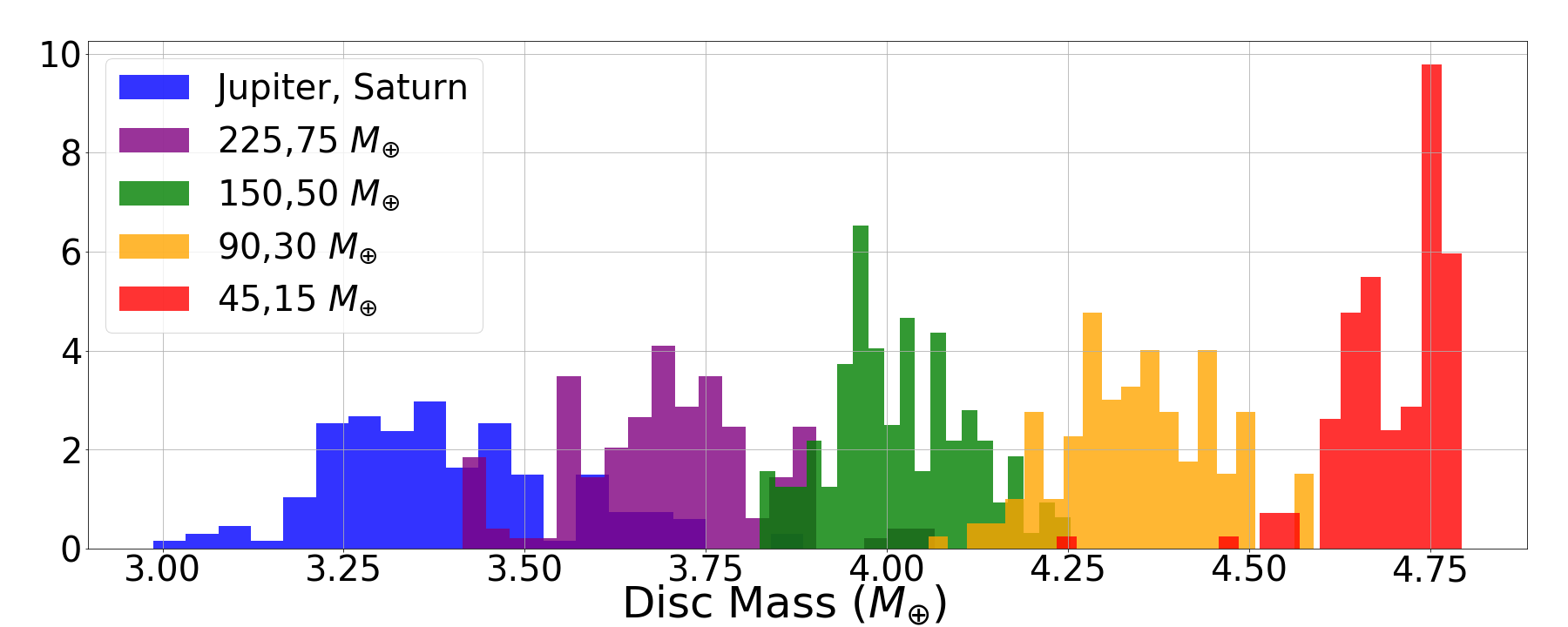}
\caption{Normalised histograms of the remaining disc mass (not including mass of the exterior giant planets) in each run at 5 Myr for each system.  Note that the more massive the giant planets, the less massive the remaining terrestrial disc.}
\label{fig:discmass}
\end{center}
\end{figure}

\begin{figure}
\begin{center}
\includegraphics[width=\columnwidth]{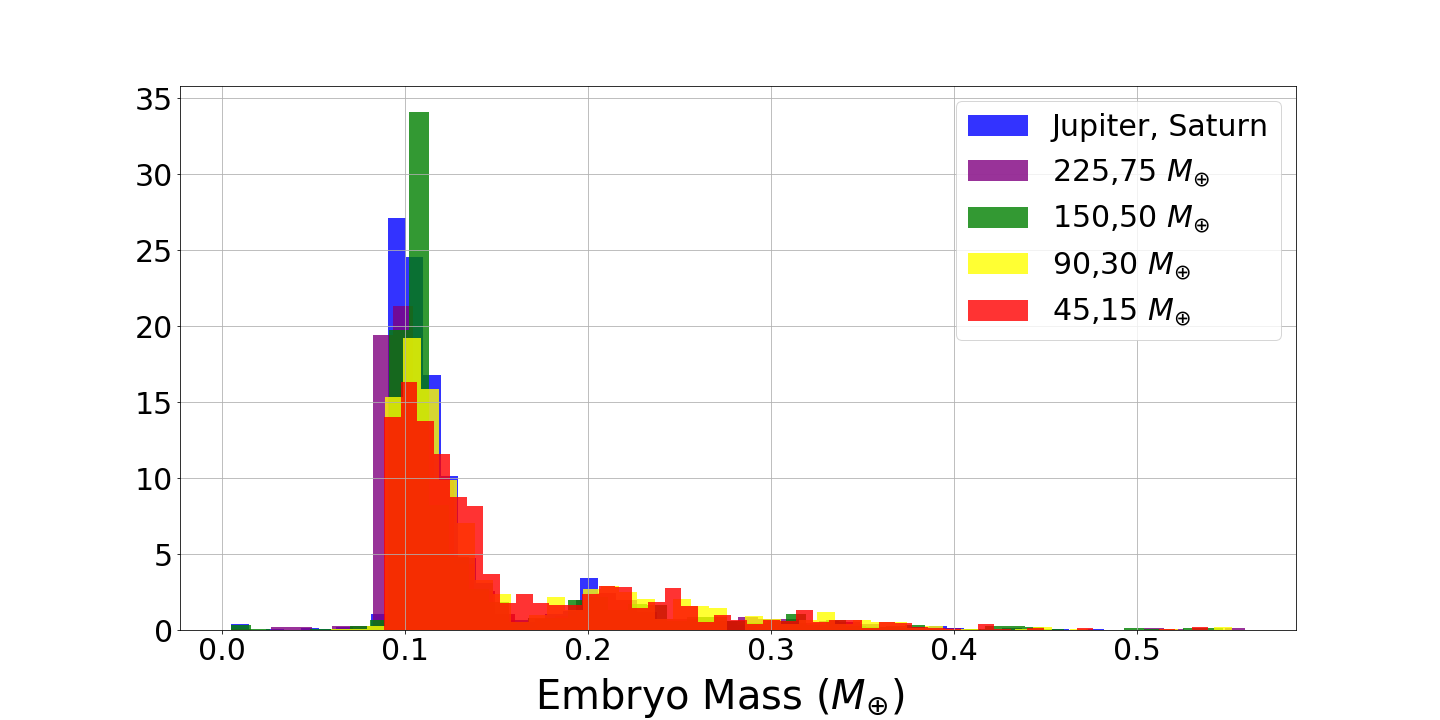}
\caption{Histogram of the remaining embryo's mass at 5 Myr for embryos with a semi-major axis less than 2 AU.  More massive giant planets tend to produce embryos with less mass at this time.}
\label{fig:2AU_embryo}
\end{center}
\end{figure}

\begin{figure}
\begin{center}
\includegraphics[width=\columnwidth]{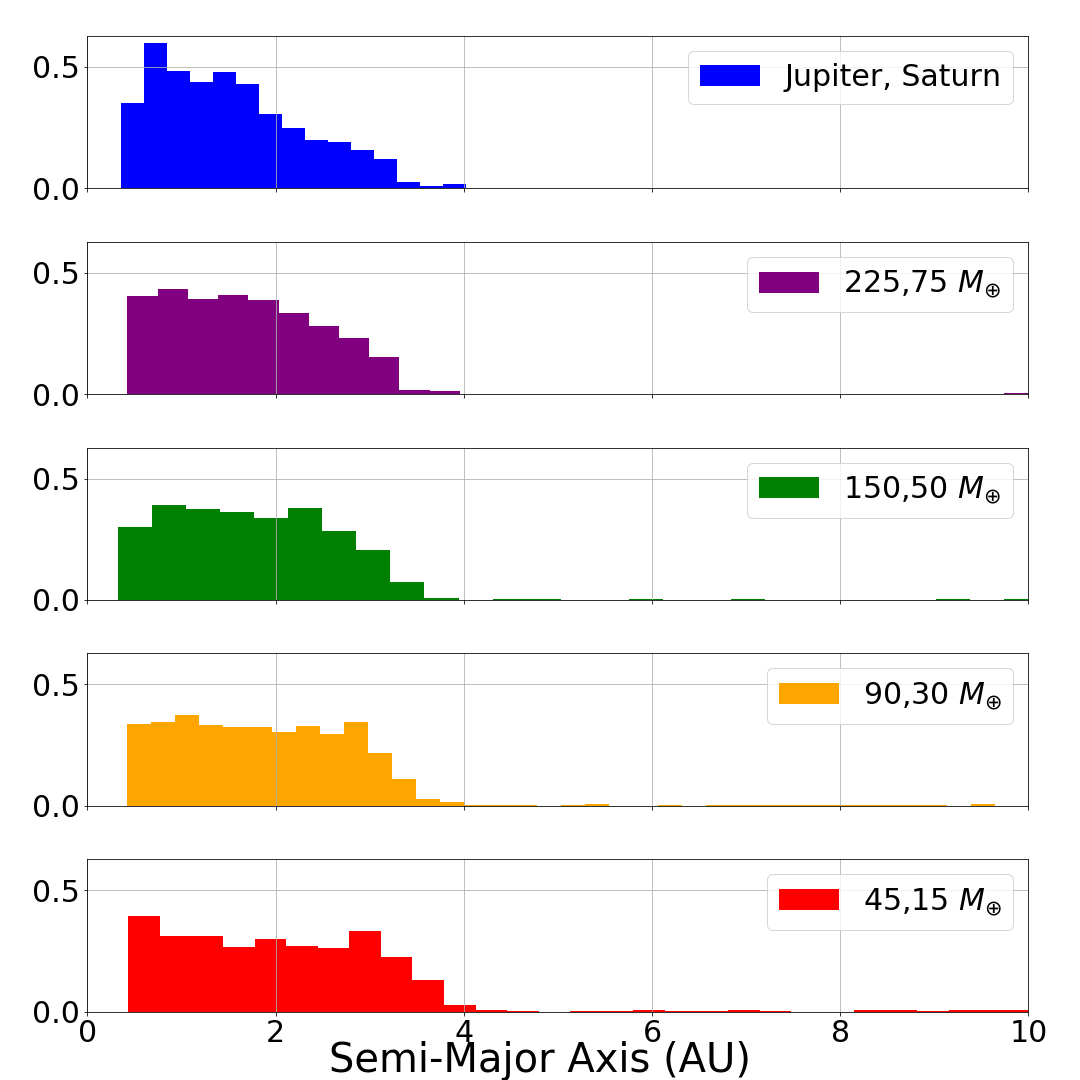}
\caption{Normalised histogram of the semi-major axis for the surviving embryos at 5 Myr.  In general, the larger the mass of the exterior giant planets, the closer in the embryos are found to the host star.}
\label{fig:semi_hist}
\end{center}
\end{figure}

\begin{figure}
\begin{center}
\includegraphics[width=\columnwidth]{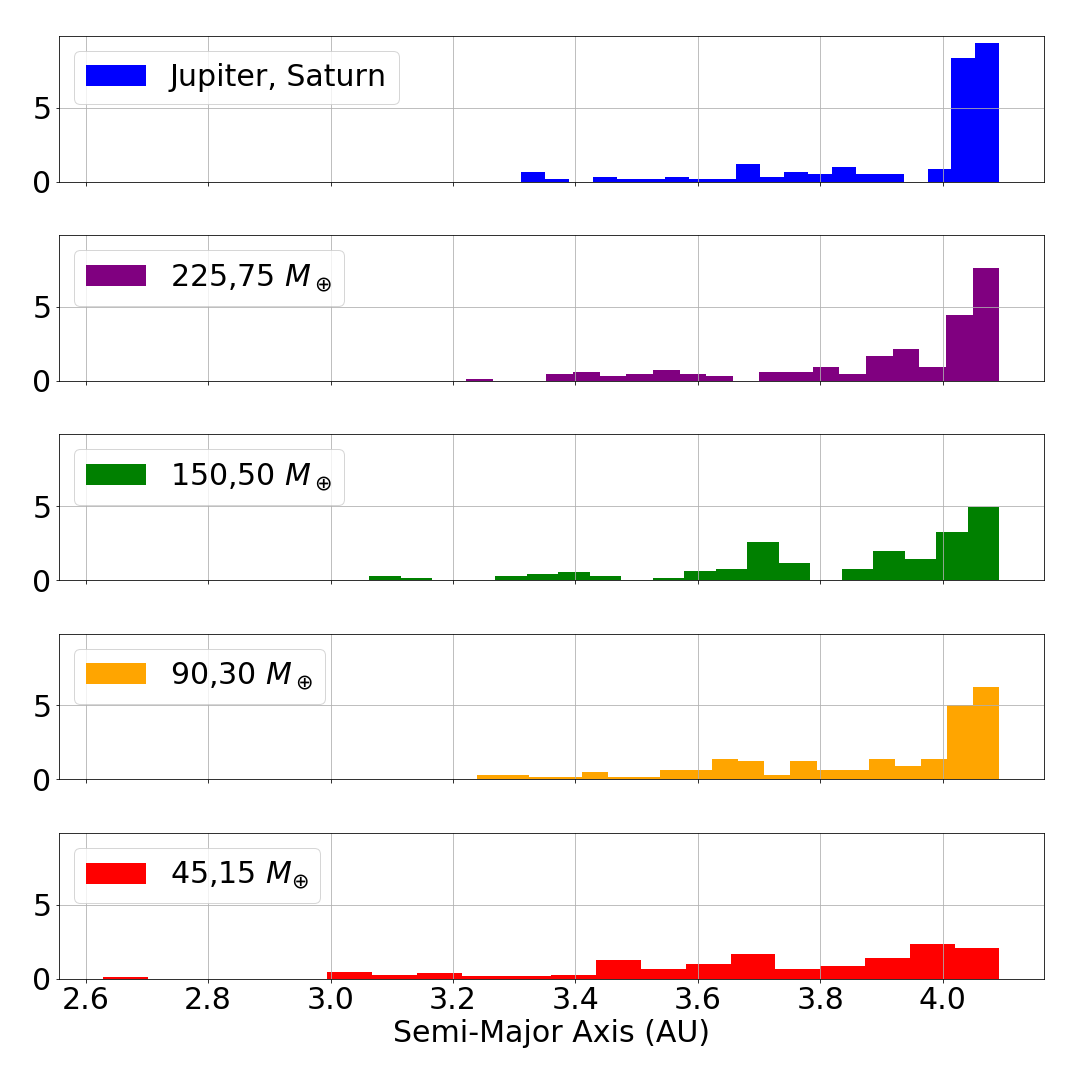}
\caption{The starting semi-major axis for all embryos that were ejected within 5 Myr.  The outer edge of the terrestrial disc is rapidly destabilized as giant planet mass increases.}
\label{fig:origins}
\end{center}
\end{figure}

\begin{figure}
\begin{center}
\includegraphics[width=\columnwidth]{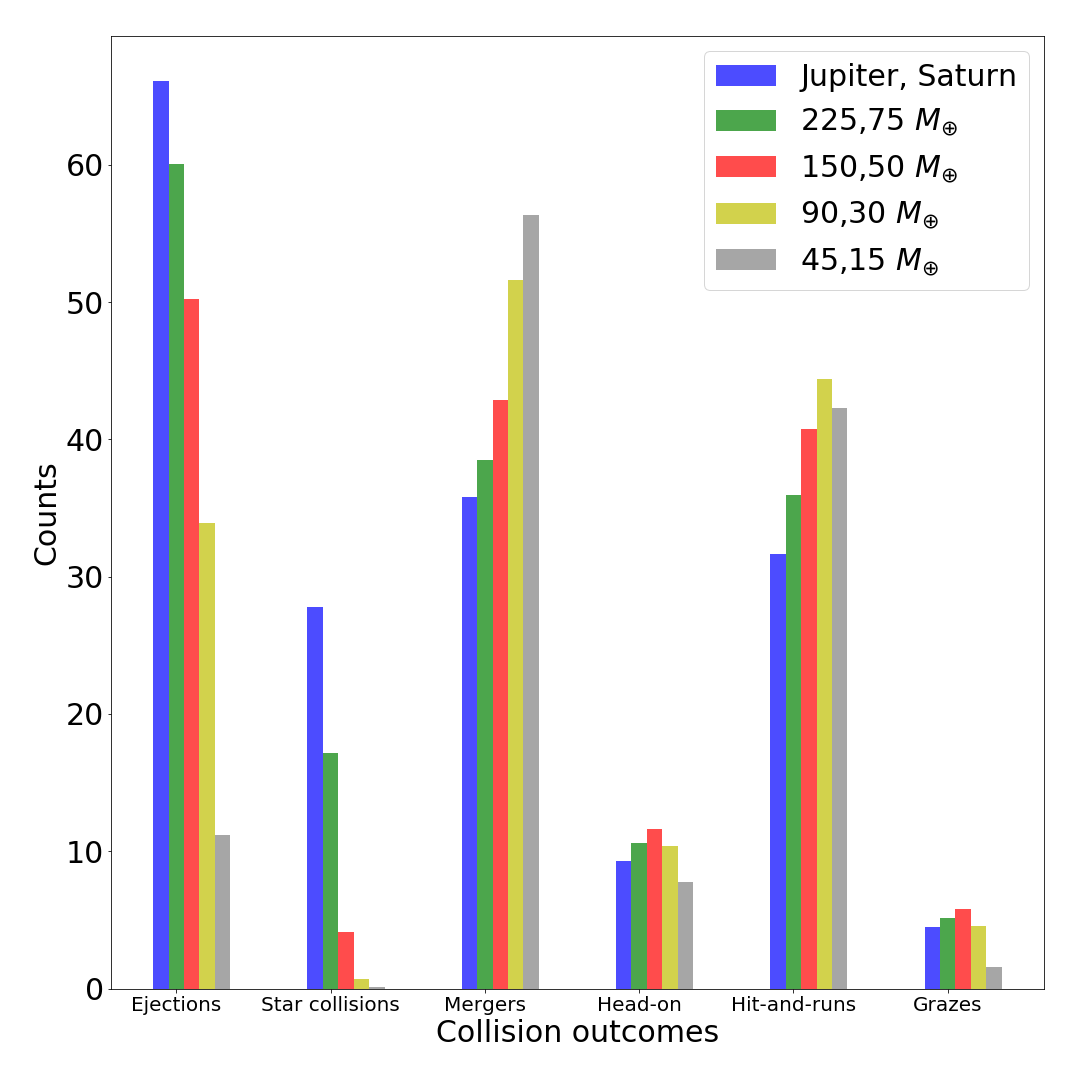}
\caption{Average counts of types of collisional outcomes after 5 Myr of integration time for all bodies in each system (i.e., embryos, planetesimals and fragments).}
\label{fig:collision_counts}
\end{center}
\end{figure}

\begin{table}
\begin{center}
\caption{The median mass, semi-major axis, and multiplicity of the embryos across all 150 simulations for each system at 5 Myr.}
\label{tab:avg_systems}
\begin{threeparttable}
\renewcommand{\arraystretch}{1.2}% for the vertical padding
\begin{tabular}{cccc}
\toprule
Exterior & Mass& Semi-major&  \# of \\ masses &  ($M_{\oplus}$)  & axis & embryos \\($M_{\oplus}$)& &(AU)& \\
\hline

Saturn \& Jupiter & 0.1 & 1.5 & 17.0\\
75 \& 225 & 0.1 & 1.6 & 18.1 \\
50 \& 150 & 0.1 & 1.8 & 18.4 \\
30 \& 90 & 0.1 & 1.9 & 19.2\\
15 \& 45  & 0.1 & 2.0 & 20.9\\

\bottomrule
\end{tabular}
\end{threeparttable}
\end{center}
\end{table}

We first consider how the planetesimal and embryo interactions evolve in time as a function of the mass of the exterior giant planets.  Figure \ref{fig:bvt} shows the total number of embryos and planetesimals in the disc versus time for our five systems.  A body ``leaves'' the system when it is ejected, as previously defined, or when it has collided with the central star.  The data for all 150 simulations were gathered together into $10^5$-year bins (this binning is used throughout the entirety of the paper)\footnote{While analysing the output of the data, we found that \textit{Mercury} with the \textit{fragmentation code} began returning erroneous results after an integration was resumed from a dump file.  After an integration was resumed from a dump file, all of the bodies were reintroduced and parsed by the \textit{element} module (which determines the orbital elements of the various objects).  As a result, there were data used by \textit{element} for bodies that had been ejected, and two data points with the same time and different orbital element values for the surviving bodies.  After learning of this bug, the data were parsed and cleaned.  If the bodies were ejected before the integration stopped, the data after the time the body was first ejected were ignored.  If the bodies survived and had two data points at the same time, the second data point was ignored.  When reviewing the corrupt data, it was obvious that the second data point was incorrect.}.  The median of each bin is the central line, and the bounds are the $16^\text{th}$ and $84^\text{th}$ percentiles. 

We see that the larger the exterior planet mass, the faster the number of bodies decreases in the system.  This finding is in agreement with \cite{Barclay17} who, using the same terrestrial disc used here, simulated systems with Jupiter and Saturn analogues and systems with no giant planets.  \cite{Barclay17} found that the systems with Saturn and Jupiter ejected one-third of the terrestrial disc within the first 25 Myr while the systems without giant planets only ejected 1\% of the terrestrial disc by the end of 2 Gyr.  Such a result is expected since a larger planet is capable of producing a larger perturbation to the orbit of the smaller bodies and more readily generates orbit crossings that lead to collisions, ejections, or accretion onto the central star \citep{Levison03}.  Table \ref{tab:reduction_rates} lists the time it takes the system to decrease the number of bodies by 10\% (plus and minus the standard deviation among all the simulations for each system).  It also lists the ratio of time it took the system to reduce the number of bodies by 10\% compared to the Jupiter \& Saturn system, and the ratio of the exterior masses to the mass of Jupiter and Saturn.

The second most massive system in our study is the system with 75 \& 225 $M_{\oplus}$ cores at Saturn's and Jupiter's orbit, respectively.  While the total exterior mass of this system is only 27\% smaller than the Jupiter+Saturn system, it takes $\sim$ 2.5 times longer to eject 10\% of the system's bodies (1.2 Myr compared to 0.5 Myr for Jupiter+Saturn).  The least massive system, 45 \& 15 $M_{\oplus}$, ejected only 7\% of the bodies in the disc before 5 Myr.  These findings show a strong relationship between exterior planet mass and the ejection rate of planetesimals and embryos. 

The ejection rate sets an upper limit on disc mass and thus the availability of material for terrestrial planets to interact with and form from.  Figure \ref{fig:discmass} shows the normalised histograms of the remaining disc mass (not including the mass from the exterior giant planets) in each run at 5 Myr for each of the systems, confirming that lower-mass giant planets retain more disc mass.

Considering the evolution of individual embryos gives insight into the types of terrestrial planets that will form.  Figure \ref{fig:2AU_embryo} shows the mass distribution of the embryos that reside within 2 AU after 5 Myr of simulation time for all 150 simulations (these are the bodies with an initial mass of $m=0.093$ $M_{\oplus}$).  For the low mass giant systems we find embryo mass to be equal to or higher than the initial embryo mass.  This result suggest that the embryos did not gain or lose a significant amount of mass and they are only accreting material.  Slightly smaller embryos are found in the Jupiter and Saturn system.  This is most likely due to the higher ejection rates which decreases the amount of material in the disc needed for embryos to grow.

The median mass, semi-major axis, and multiplicity for the remaining embryos in each of the systems at this time is summarised in Table \ref{tab:avg_systems}.  Although the median mass of an embryo is the same for each system at this time, we find an anti-correlation between semi-major axis of the embryo and exterior giant planet mass, and also the number of embryos and the exterior giant planet mass as seen in Figure \ref{fig:semi_hist}.  We find that more massive giant planets produce embryo systems closer to their host star because the giant planets are more likely to eject material from the outer edge of the terrestrial disc.  Figure \ref{fig:origins} shows the starting semi-major axis for all the embryos and planetesimals ejected within the first 5 Myr of simulation time. As giant planet mass increases, it rapidly destabilizes the outer part of the terrestrial disc.  This finding explains the higher ejection rates, and smaller disc mass and semi-major axis distributions for the terrestrial protoplanets in the simulations with more massive giants.

Using the \textit{fragmentation code}, we are able to follow the collision history of the embryos in higher resolution than done in previous numerical studies.  The collision history of a terrestrial system can affect the composition/volatile budget \citep{Horner09}, atmosphere, internal structure, rotation, moon formation or moon-forming debris \citep{Kegerreis18,Citron18}).  Figure \ref{fig:collision_counts} shows the average occurrence rate for types of collision outcome in each system after 5 Myr of simulation time.  These are the collision rates for all embryos, planetesimals and fragments across all 150 runs for each system.  Because massive giant planets are more likely to eject disc mass and produce collisions with the central star, we see a higher rate of mergers and hit-and-run collisions in smaller giant planet systems that have retained more mass.  We find a similar distribution between head-on and grazing collisions between all systems.

A head-on collision will not necessarily result in a super-catastrophic collision. If the specific impact energy is too low, a super-catastrophic collision will not take place and fragments will not be produced.  Embryos that do not have an extensive history of catastrophic collisions will be able to grow their planet size more readily and maintain a thicker atmosphere, but their likelihood of forming a moon system or magnetic field decreases \citep{Jacobson17}.  We have not seen any definitive moon systems in the Kepler data, which are predominantly planets of several Earth masses and sizable atmospheres---though Kepler data may not be sufficiently sensitive for most small moons.  \cite{Teachey18} however, found evidence from the Hubble Space Telescope that suggests the Jupiter sized planet Kepler-1625b may be associated with a Neptune sized moon.

\begin{figure}
\begin{center}
\includegraphics[width=\columnwidth]{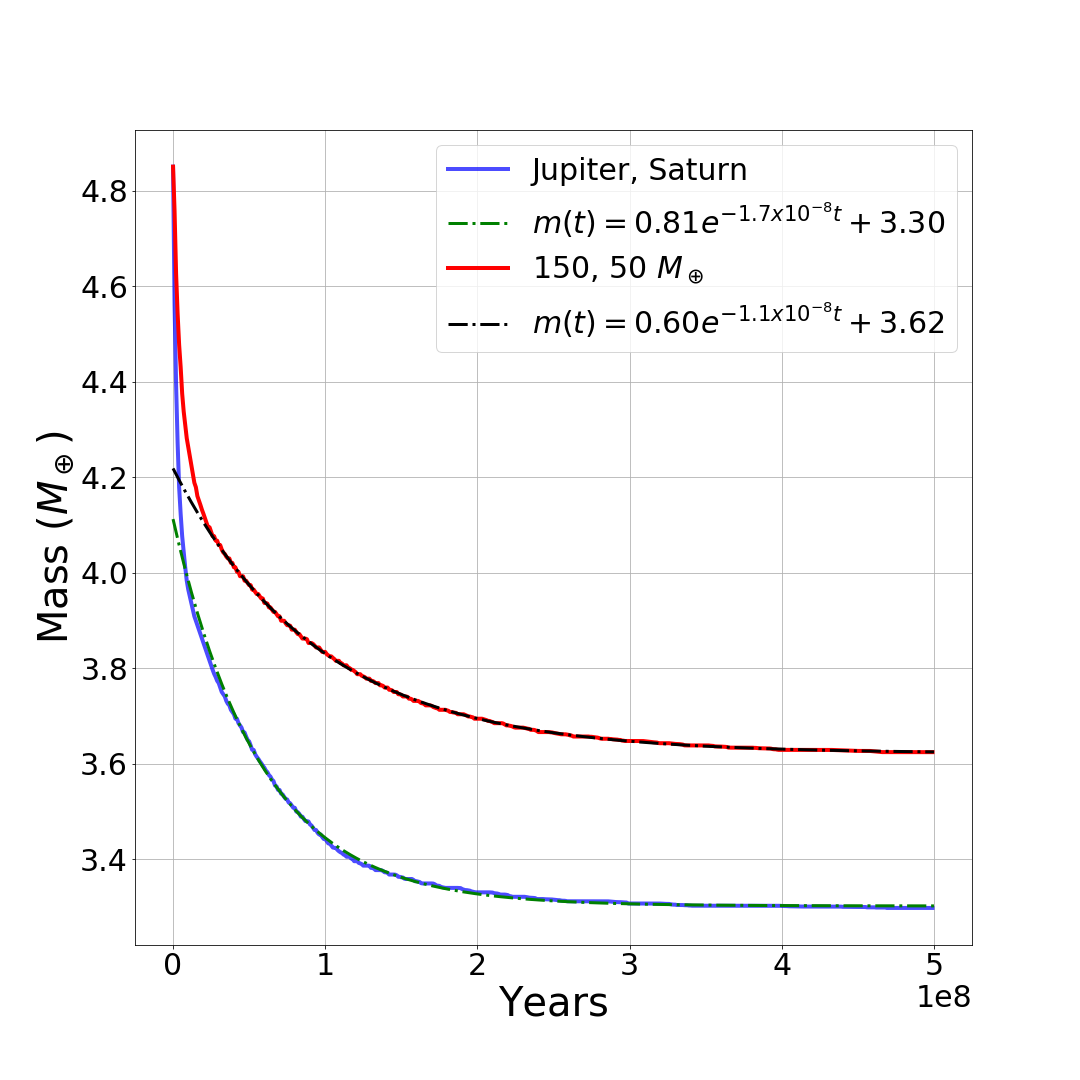}
\caption{Median disc mass versus time of all 150 realisations for the Jupiter \& Saturn system, and the 150 \& 50$M_{\oplus}$ system.  Decaying exponential fits to the data beyond 2 Myr are shown for each system.  Note that after a few hundred million years, the disc mass remains roughly constant with the lower-mass giants producing a more massive disc.}
\label{fig:exp_fits}
\end{center}
\end{figure}

\begin{figure}
\begin{center}
\includegraphics[width=\columnwidth]{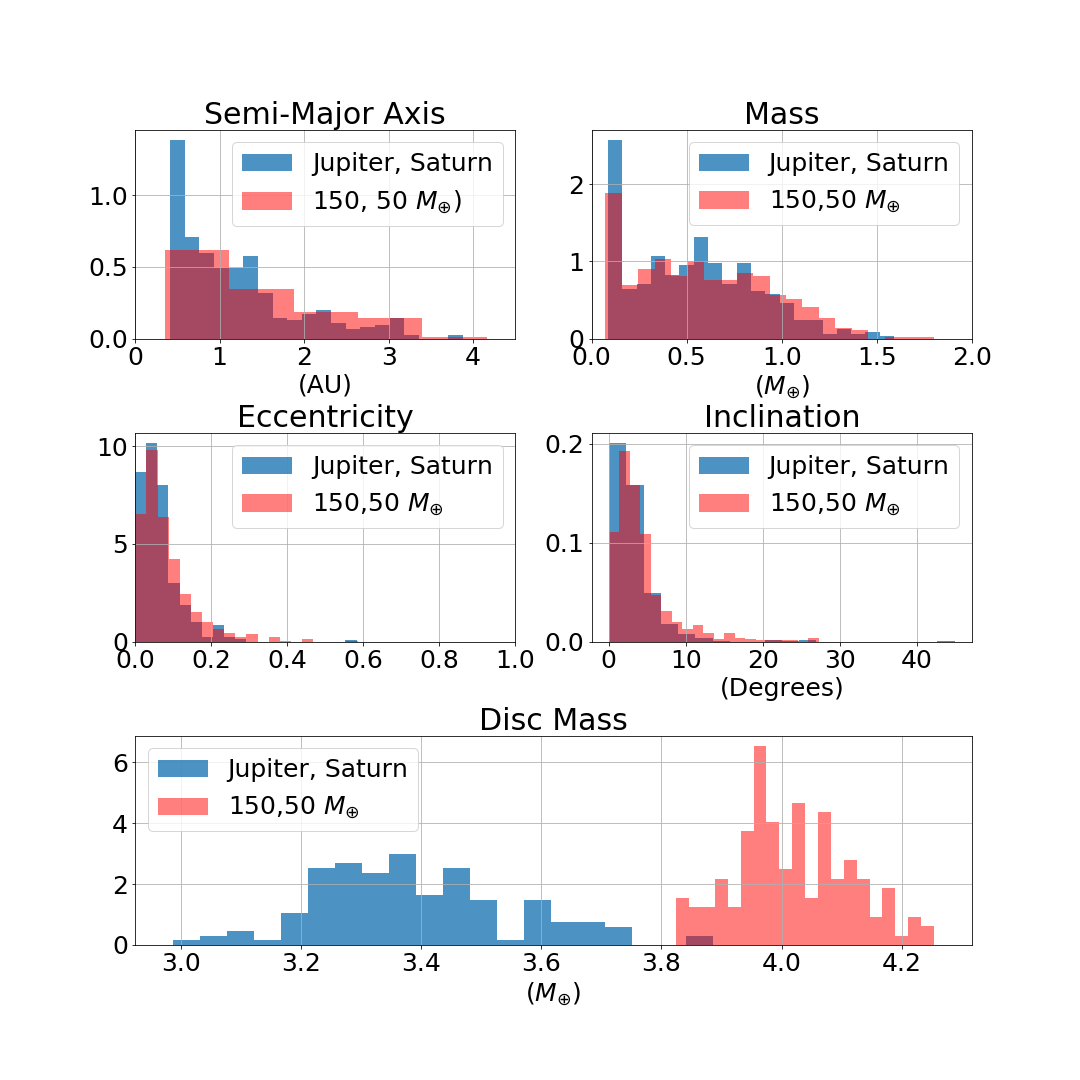}
\caption{Normalised histograms showing the distribution of the mass, semi-major axis, eccentricity and inclination for the surviving embryos and the total disc mass at $\sim$ 500 Myr.  We see an anti-correlation between disc mass and exterior giant planet size.}
\label{fig:ext_semi_mass_hist}
\end{center}
\end{figure}

\begin{table}
\begin{center}
\caption{Median values and standard deviations of the embryo orbital elements at 500 Myr.
}
\label{tab:embryo_params}
\begin{threeparttable}
\renewcommand{\arraystretch}{1.2}% for the vertical padding
\begin{tabular}{cccc}
\toprule
\textbf{Jupiter, Saturn system} \\
\hline
 Orbital Element & Median Value & Standard Deviation\\
\hline
Semi-Major Axis (AU) & 0.9 & 1.2 \\
Mass ($M_{\oplus}$) & 0.54 & 0.33 \\
Eccentricity &0.05 & 0.06 \\
Inclination (Degrees) & 2.4 & 4.1 \\
\hline
\textbf{150, 50 $M_{\oplus}$ System} \\
\hline
Semi-Major Axis (AU) & 1.1 & 1.1\\
Mass ($M_{\oplus}$)& 0.57 & 0.36 \\
Eccentricity &0.06 & 0.07\\
Inclination (Degrees) & 3.0 & 3.6\\
\bottomrule
\end{tabular}
\end{threeparttable}
\end{center}
\end{table}

\begin{figure}
\begin{center}
\includegraphics[width=\columnwidth]{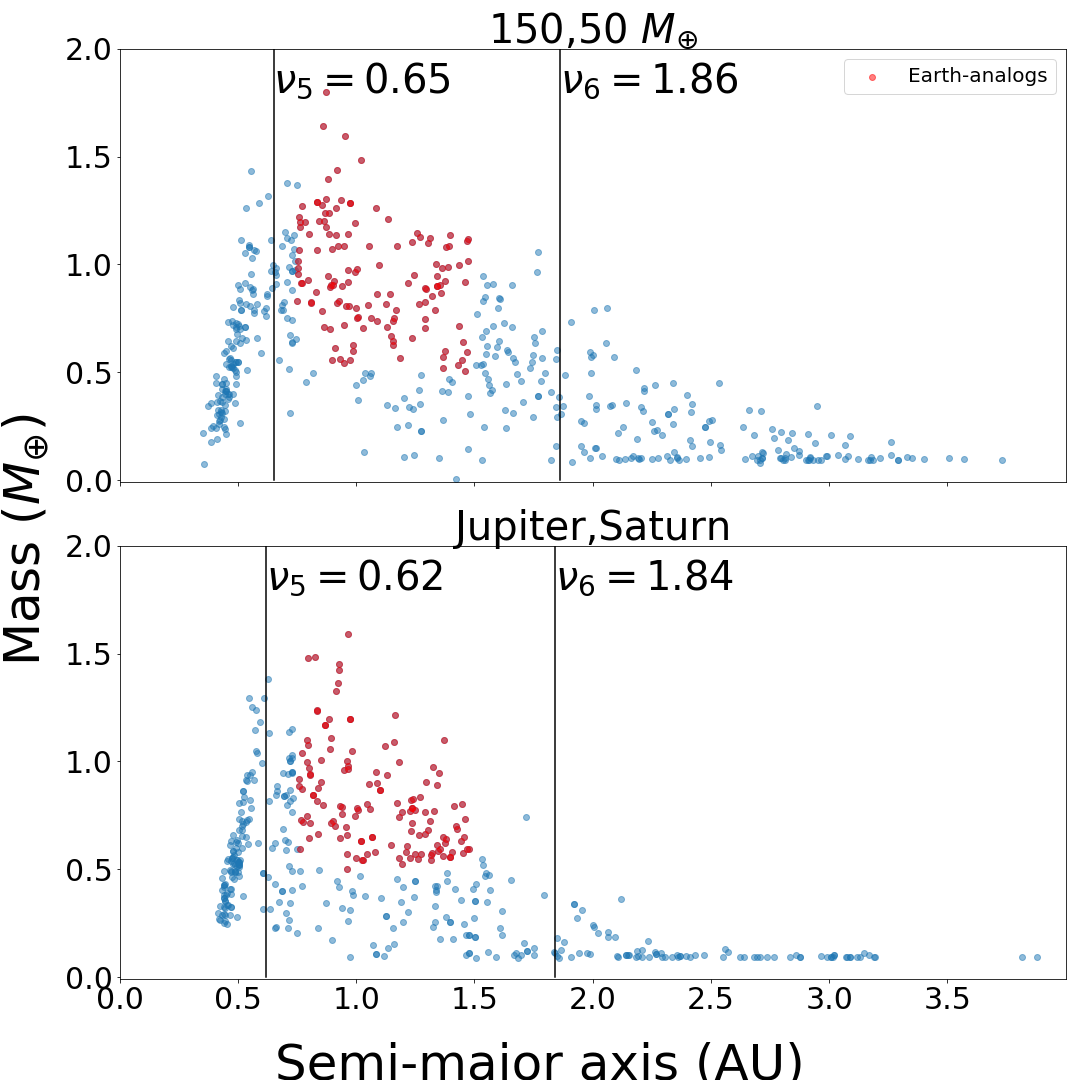}
\caption{Remaining embryo mass ($M_{\oplus}$) and semi-major axis (AU) for the Jupiter \& Saturn, and 150 \& 50$M_{\oplus}$ systems after $\sim$ 500 Myr.  The $\nu_5$ secular resonance induced by the planet at Jupiter's orbit and $\nu_6$ secular resonance induced by the planet at Saturn's orbit are shown in solid black lines.  The red dots are the Earth analogues.  We find a similar distribution of Earth analogues between the two systems.}
\label{fig:ext_system_structure}
\end{center}
\end{figure}

\begin{figure}
\begin{center}
\includegraphics[width=\columnwidth]{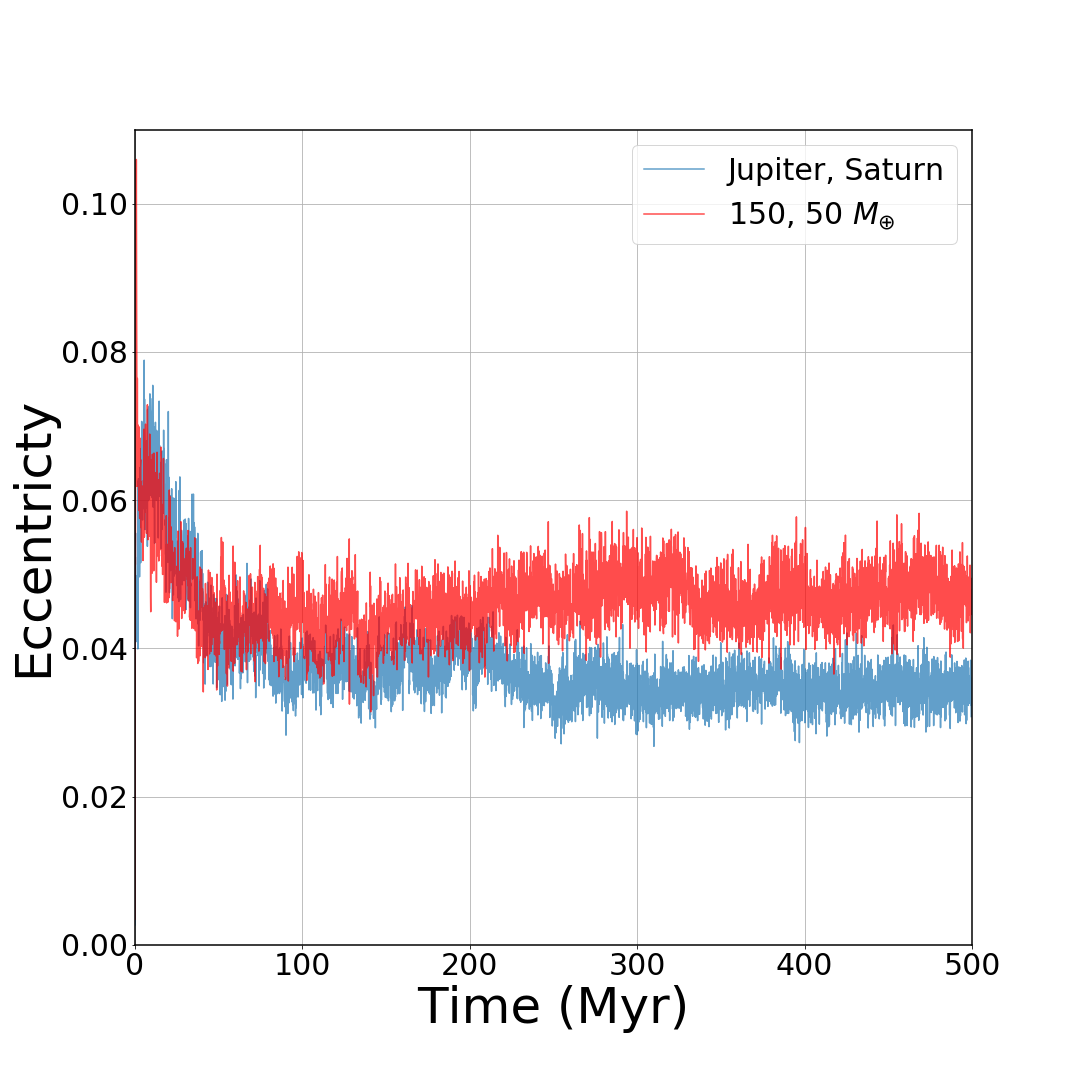}
\caption{Median eccentricity versus time for all Earth analogues at 500 Myr.  The Jupiter and Saturn system dampen the eccentricities more than the system with $\sim \tfrac{1}{2}$ the mass at Saturn's and Jupiter's orbit, however both systems show a similar median eccentricity for Earth analogues.}
\label{fig:ext_ecc}
\end{center}
\end{figure}

\begin{figure}
\begin{center}
\includegraphics[width=\columnwidth]{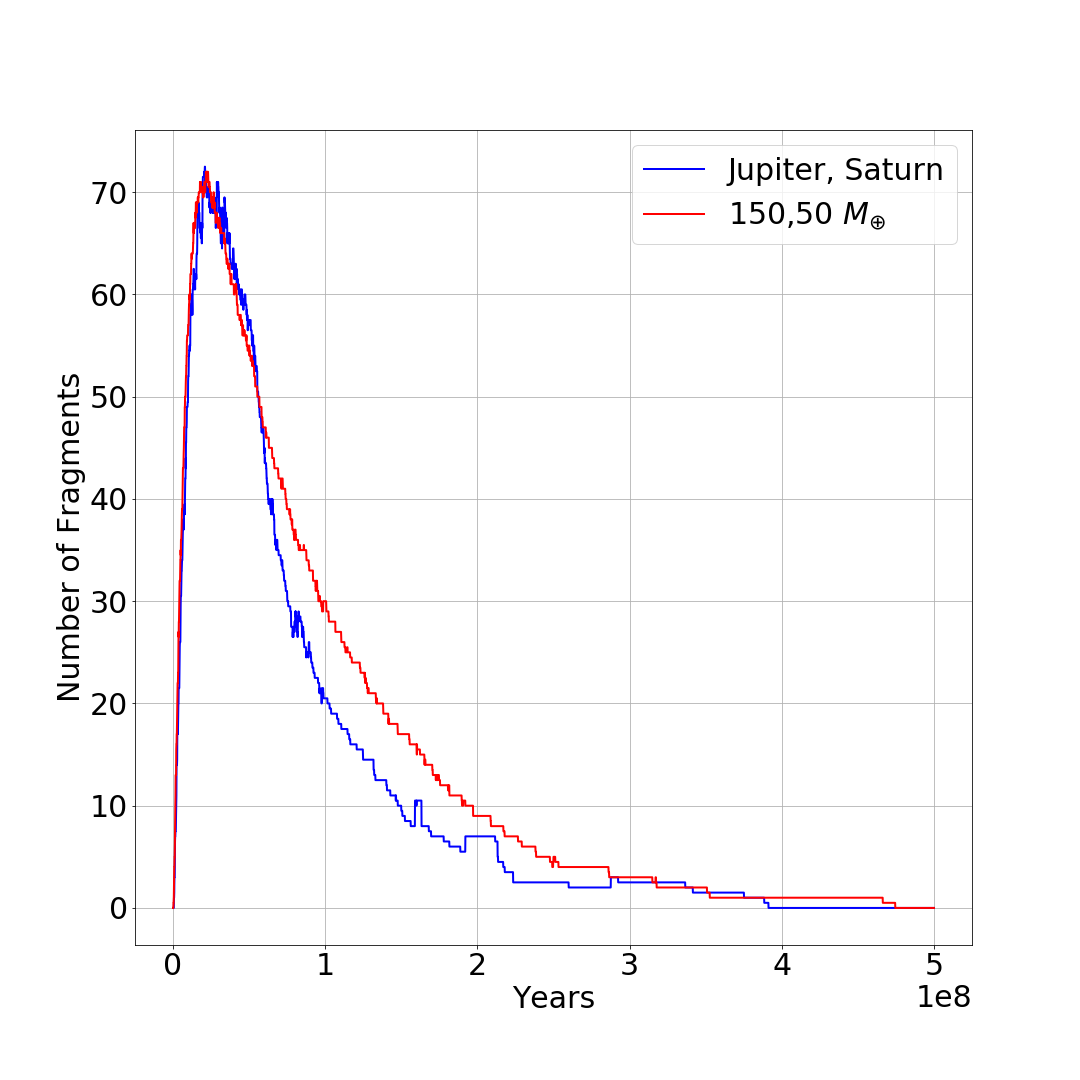}
\caption{Median number of fragments in the system versus time for our two extended systems.  The system with the more massive exterior giant planets produces fewer fragments after $\approx$ 50 Myr.  Both systems have few fragments after 400 Myr.}
\label{fig:frag_counts}
\end{center}
\end{figure}

\begin{figure}
\begin{center}
\includegraphics[width=\columnwidth]{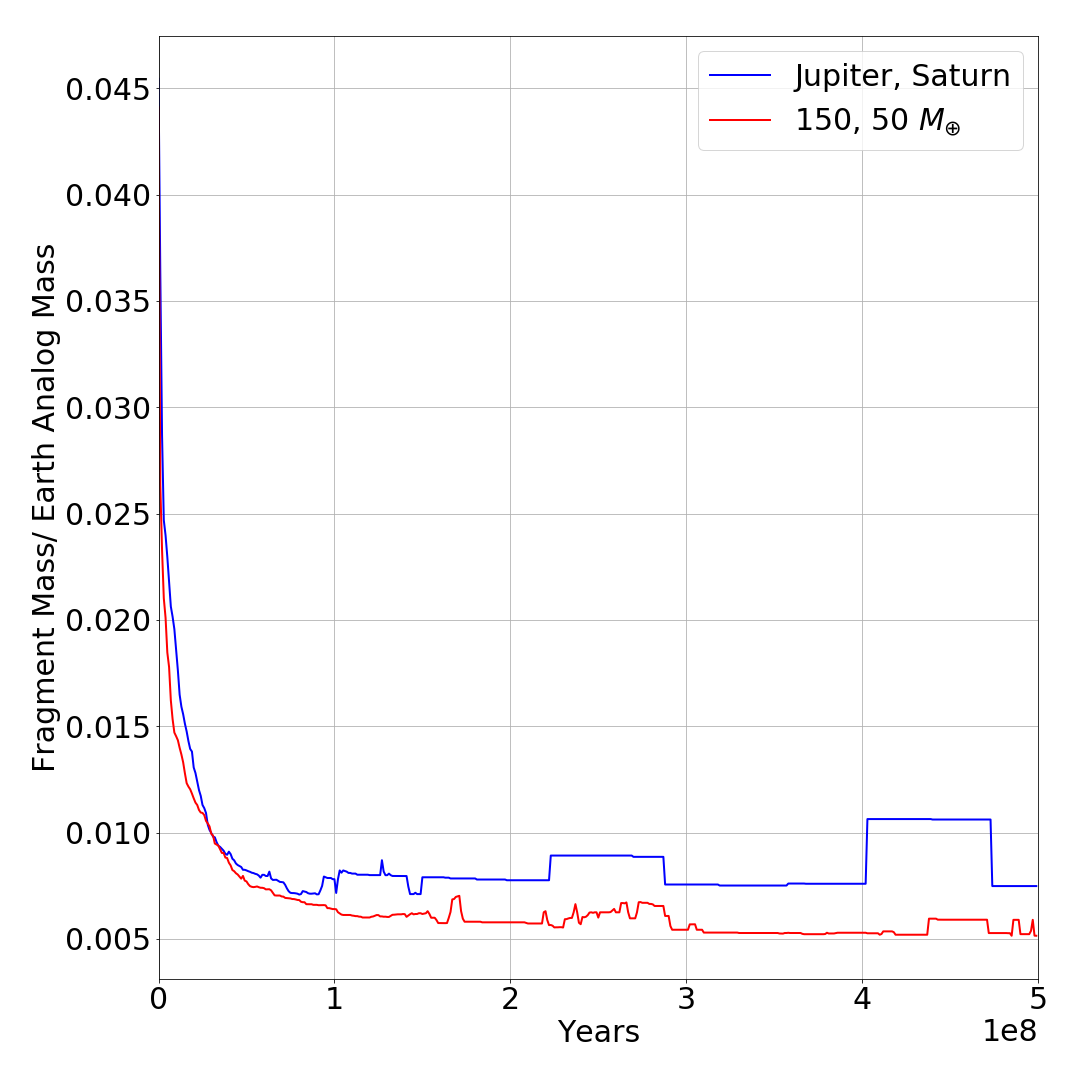}
\caption{Ratio of median fragment mass to median Earth analog mass versus time.  This ratio becomes larger in the Jupiter and Saturn system around the same time the Earth analog eccentricities become more dampened than those in the 150, 50 $M_{\oplus}$ system ($\approx$ 50 Myr).}
\label{fig:frag_mass}
\end{center}
\end{figure}

\section{Extended Runs}

For a better understanding of what the final terrestrial system will look like, we integrate the systems with Jupiter \& Saturn and 150 \& 50$M_{\oplus}$ exterior planets up to 500 Myr.  Figure \ref{fig:exp_fits} shows the mass versus time for all 150 simulations of the two extended runs.  The mass loss rate begins to slow down and flatten out before 200 Myr. The decrease in slope suggests that the reservoir of planet forming material is roughly constant from that point onward.  The mass loss rate associated within the first Myr of the system is primarily a manifestation of the sudden onset of gravity rather than the ongoing interactions with the exterior giants.  Because of these effects, we fit the mass loss rate curves after 50 Myr, where the fits become more constant than the fits done at earlier times.  The fits of these curves follow the form,
\begin{equation}
m(t) = A \exp\left[- \tfrac{t}{\tau}\right]+C.
\end{equation}
where $\tau$ is the relaxation timescale for the system---a measure of how long the disc will be affected by the perturbations from the exterior giant planets.  For the Jupiter \& Saturn system $ \tau_5 \approx$ 58 Myr, and for the 150 \& 50 $M_{\oplus}$ system $\tau_6 \approx$ 95 Myr showing an inverse relationship between giant planet mass and the relaxation timescale.  As a more massive planet produces stronger perturbations to the disc, the scattering will occur more quickly and reach a constant disc mass sooner.

Again, we consider the orbital elements and mass of the embryos at 500 Myr so we may get an idea of the properties of the final terrestrial planet system.  Figure \ref{fig:ext_semi_mass_hist} shows the normalised histograms of the mass, semi-major axis, eccentricity and inclination for the surviving embryos at $\sim$ 500 Myr, and the remaining disc mass in each run.  The median remaining disc mass at this time is $\sim$ 3.30$ M_{\oplus}$ for the Jupiter \& Saturn system, and $\sim$ 3.62$M_{\oplus}$ for the 150 \& 50 $M_{\oplus}$ which suggests an anti-correlation between exterior giant planet mass and disc mass.  Table \ref{tab:embryo_params} lists the median values and standard deviations of the embryos orbital elements at 500 Myr.  While both systems have similar embryo mass and orbital distributions, the Jupiter \& Saturn systems tend to produce embryos that are lower in semi-major axis, mass, eccentricity, and inclination.  However, the largest fractional difference between the two systems is in the inclination values, which is only $\approx$ 0.20 with the 150, 50 $M_{\oplus}$ system producing more inclined embryos.  Again, the lower semi-major axis, embryo mass, and embryo multiplicity in the Jupiter and Saturn system may be a result of a higher ejection rate of terrestrial material from the outer edge of the terrestrial disc.

Another feature that we see in these extended runs relates to the secular dynamics of the system.  Figure \ref{fig:ext_system_structure} shows the mass versus semi-major axis distribution of the embryos in the 150, 50 $M_{\oplus}$ and Jupiter, Saturn systems after $\sim$ 500 Myr with the locations of the secular resonances from Jupiter and Saturn marked by $\nu_5$ and $\nu_6$, respectively, found with methods of \cite{Smallwood18}.  What we find is a similar mass and semi-major axis distribution of the embryos between the two systems.  Interestingly, the highest mass embryos are on orbits with precession rates that lie near to, or between the secular frequencies of the giant planets.  There is a particularly strong feature just interior to the secular frequency of the largest giant.  Upon closer inspection of the giant planet orbital evolution, we find that both giant planets move inwards in the beginning of all simulations.  As the giant planets scatter material out of the system, their secular resonances drift inwards.  As the orbits of the surviving bodies evolve, many are shepherded by this resonance, producing an overabundance of planet-forming material in its vicinity.  The gap in the semi-major axis distribution at the locations of the secular resonances is due to instability, and eventual ejection, caused by orbital chaos at the location of these resonances.  This result may have observable consequences in exoplanetary systems---it may be that the most massive terrestrial planets are more likely to form in the regions just interior to the locations of strong secular resonances.  This result is confirmed by the findings of \cite{Hoffmann17} who also used numerical studies to show that the mass distribution peaks at the $\nu_5$ resonance, and the outer edge of the terrestrial system is marked by the $\nu_6$ resonance.

We define an Earth analog, using criteria from \cite{Quintana16}, as a planet found between 0.75-1.5 AU and a mass $>$ 0.5 $M_{\oplus}$.  The system with 150 \& 50 $M_\oplus$ planets at Jupiter's and Saturn's orbit is just as efficient at creating Earth analogues as the system with Jupiter and Saturn---with a total of 131 and 126 Earth analogues respectively at 500 Myr.  On average, both giant planet systems produced one Earth analog per run.  The total number of Earth analogues produced in the Jupiter \& Saturn and 150 \& 50 $M_{\oplus}$ runs at 500 Myr are shown in red in Figure \ref{fig:ext_system_structure}.  The similar distributions of the Earth analogues between the two systems suggests that a fairly wide range of giant planet masses can produce Earth analogues.

As the Milankovitch cycles provide climatic variability on Earth, a moderate eccentric orbit is needed to induce the climate changes required by evolving life \citep{Hays76}.  \cite{Horner15} found that changes to Jupiter's orbit can have significant effects on Earth's eccentricity, and thus climate.  If the eccentricity of Earth's orbit becomes too high, it can have catastrophic effects for life.  Consequently, we examine the eccentricity evolution of the embryo's that become Earth analogues at 500 Myr to predict if the resulting planets will have a habitable eccentricity.  The trend in embryo eccentricity is seen in Figure \ref{fig:ext_ecc}, which shows the median eccentricity versus time for all of the surviving embryos in the two systems that were integrated for 500 Myr.  Although the Jupiter and Saturn system yield planet eccentricities that are dampened more than the system with the smaller giant planets, the two systems show a similar eccentricity evolution.  Both systems produce Earth analogues with an eccentricity between 0.03 and 0.06 after 500 Myr, which is larger than modern Earth's eccentricity.

While the eccentricity differences between the two systems are small, the trend is unexpected since we would expect to find more excited eccentricities in a system with less material and fewer bodies \citep{Obrien06}.  Figure \ref{fig:frag_counts} shows the median number of fragments in our two extended systems versus time.  The more massive giant planets produce fewer, but larger fragments after 100 Myr.  Figure \ref{fig:frag_mass} shows the ratio of the median fragment mass to median Earth analog mass.  This fact suggests that the Earth analog eccentricities are not being damped by dynamical friction.

\section{Conclusions}

We set out to understand some of the ways that exterior giant planets mold the interior terrestrial planet system.  We consider five different systems, each with scaled-down masses of Saturn and Jupiter at Saturn's and Jupiter's orbit.  We ran 150 simulations of each system with slight variations to the disc, for at least 5 Myr.  We use an updated \textit{fragmentation code} to model collisions between bodies in higher resolution than previous N-body studies.  Our results show that exterior giant planet mass does affect some aspects of the architecture of the terrestrial system---its formation timescale, and the details of the terrestrial planet's collision history.  Interestingly, we do not find that changing the giant planet mass at Jupiter's and Saturn's orbit significantly affects the number of Earth analogues (a planet found between 0.75-1.5 AU with a mass $>$ 0.5 $M_{\oplus}$) produced in our simulations.

Because the maximum integration time is only 500 Myr, we consider the evolution of the embryos and make statements about the expected properties of the resulting planets.  The embryo's in the systems containing lower-mass giant planets are more likely to experience accretion events in the form of perfect mergers, and systems containing higher mass exterior giant planets are more likely to eject material from the outer edge of the terrestrial disc. 

We compare the collision rates among all the systems used in our simulations after 5 Myr.  The largest statistical difference is found in the ejections rates.  Simulations with larger giant planets eject more material.  For example, the Jupiter and Saturn system ejects more than six times the amount of matter than the system with the lowest mass giant planets (45 \& 15 $M_{\oplus}$).  Further analysis shows that the larger the mass of the exterior giant planet, the more likely the ejections originate from the outer edge of the terrestrial disc.  This truncation of the outer terrestrial disc leads to terrestrial planets on shorter orbits.
 
Among all of the systems, with the exception of the system with the smallest giant planet masses, the grazing events are similar.  The grazing rates may affect the density of the final planets as grazing impacts tend to remove the outer portion of the embryo.  If the embryo is differentiated at this time, a grazing event will increase the density of the body.  The system with 45 \& 15 $M_{\oplus}$ exterior giant planets has the lowest rate of grazing impacts with less than half the number of grazing events than the Jupiter and Saturn system.  This suggests that systems with smaller exterior giant planets may produce terrestrial planets with a lower density.  All systems have a similar rate for head on collisions, but Jupiter and Saturn produce only $\tfrac{2}{3}$ the amount of mergers and $\tfrac{3}{4}$ the amount of hit-and-run interactions than the system with the smallest exterior planets.  This occurs because Jupiter and Saturn have a higher ejection rate and therefore less material in the terrestrial disc than the other systems.

The eccentricity of a planet has implications for the habitability of the planet.  A mildly eccentric orbit on Earth is needed to provide the climatic variability needed for the evolution of life (the Milankovitch cycles)\citep{Hays76}, however a highly eccentric orbit can have catastrophic consequences for life \citep{Hornerhabit10}.  Our simulations show that the larger exterior giant planets induce lower eccentricity orbits after $\sim 10$ Myr for the Earth analogues in the system.  We also find that systems with more massive giant planets produce fewer fragments late in the evolution of the system, but a larger median fragment mass to median Earth analog mass ratio.

Again, more massive giant planets yield terrestrial planet systems that are closer to the central star given our initial disc, and the secular resonances they produce affect the final locations of the terrestrial planets.  As the giant planets scatter the planet-forming material, the locations of the secular resonances move inward---shepherding material as they sweep by.  Interior to both resonances we find larger embryos, with the largest ones near the resonances.  At the same time, the chaotic regions produced near these secular resonances prevent the formation of terrestrial planets that are too close to the resonance locations.

Placing constraints on the time needed for terrestrial planets to form is essential for a complete understanding of a general planet formation theory.  Our results show that the relaxation timescale (the time in which the giant planets significantly perturb the terrestrial material) is sensitive to exterior giant planet mass.  There exists an inverse correlation between exterior giant planet mass and this relaxation timescale, $\tau$.  This relationship results from the stronger perturbations caused by more massive planets.  Because smaller exterior giant planets interact with the disc for a longer period of time than the larger giant planets, we expect the later system to reach a dynamically stable configuration earlier in time.

Future exoplanet missions, such as WFIRST, will be sensitive to planets on longer orbital periods compared to those found by NASA's Kepler mission, or those expected from the recently launched Transiting Exoplanet Survey Satellite (TESS).  With the launch of WFIRST we hope to measure the occurrence rate for giant planets on longer orbits, and how they correlate with interior terrestrial planet systems.  Such a measurement will determine how common solar system analogues are in the galaxy.

\section*{Acknowledgements}
We would like to thank the anonymous referee for detailed comments that led to a much improved manuscript.  We also thank John Chambers for allowing us to use his collision fragmentation code and the UNLV National Supercomputing Institute for allowing us to use their resources for this study.  ACC and JHS are supported by NASA grants NNX16AK32G and NNX16AK08G.

\bibliographystyle{mnras}
\bibliography{references}

\label{lastpage}

\end{document}